\documentclass[a4paper,11pt]{article}
\usepackage{pos}
\usepackage{graphicx}
\usepackage{graphbox}
\usepackage{hyperref}
\newcommand{\abs}[1]{\left| #1 \right|}
\newcommand{\mc}[1]{\mathcal{#1}}

\title{Neutral meson mixing in the $B^0_{(s)}$ sector from Lattice QCD}

\author*[a]{Justus Tobias Tsang}

\affiliation[a]{IMADA \& CP$^3$-Origins. University of Southern Denmark.
  Campusvej 55, DK-5230 Odense, Denmark}

\emailAdd{tsang@imada.sdu.dk}

\abstract{We summarise the current status of lattice QCD computations of neutral
  meson parameters in the $B^0_d$ and the $B^0_s$ systems. We comment on recent
  results and anticipate further improvements on the uncertainties in the
  future.}

\FullConference{
  11th International Workshop on the CKM Unitarity Triangle (CKM2021)\\
  22-26 November 2021\\
  The University of Melbourne, Australia
}

\begin{document}
\maketitle

\section{Introduction}
\begin{figure}
  \begin{minipage}{6in}
    \centering
    \includegraphics[align=c,width=.47\textwidth]{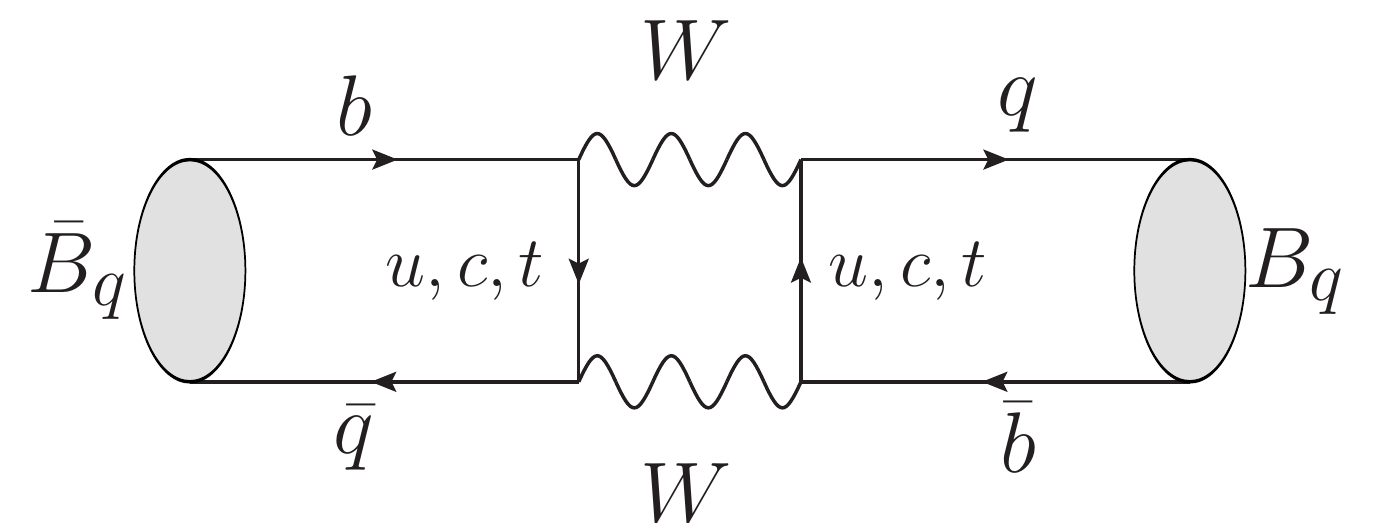}
    \hspace*{.2in}
    \includegraphics[align=c,width=.47\textwidth]{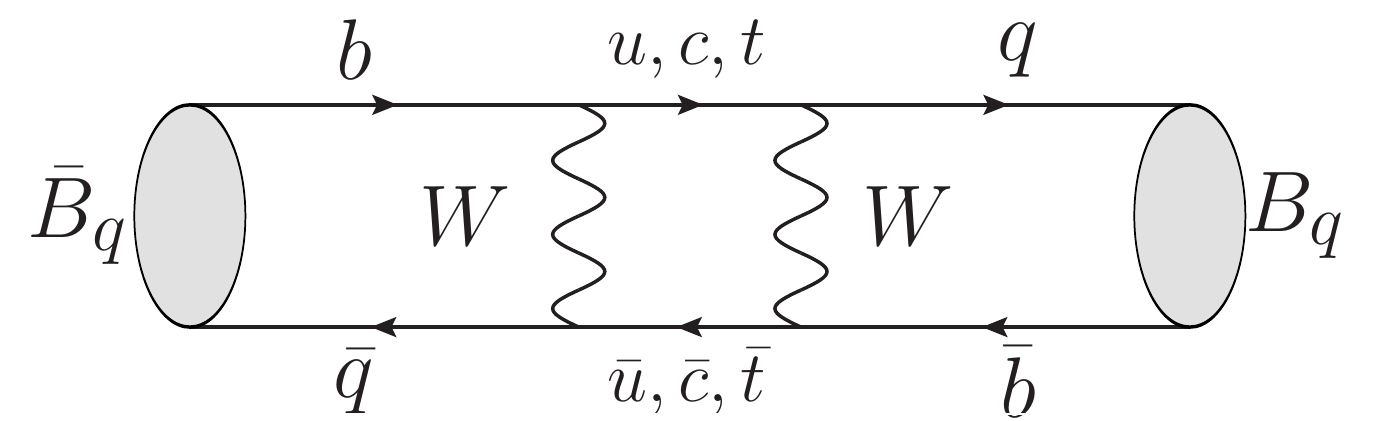}
  \end{minipage}
  \caption{Quark flow diagrams of the box diagrams mediating neutral meson
    mixing for the case of $B^0_{(s)}-\overline{B}^0_{(s)}$ mesons.}
  \label{fig:boxdiags}
\end{figure}

The flavour eigenstates of neutral mesons oscillate. This yields an
experimentally precisely measurable mass and width difference between the CP
eigenstates of the $B^0_{(s)}$-$\overline{B}_{(s)}^0$-system. The most recent
experimental measurements for the mass differences $\Delta m_d^\mathrm{exp}$ and
$\Delta m_s^\mathrm{exp}$ for the $B_d-\overline{B}_d$ and the $B_s-\overline{B}_s$
systems, respectively are~\cite{LHCb:2021moh,Amhis:2019ckw}
\begin{equation}
  \begin{aligned}
    \Delta m^\mathrm{exp}_d &= (0.5065 \pm 0.0019) \,\mathrm{ps}^{-1}\,, \\
    \Delta m^\mathrm{exp}_s & = (17.7656 \pm 0.0057) \,\mathrm{ps}^{-1}\,. \\
  \end{aligned}
\end{equation}
These oscillations are mediated through the quark flow box diagrams displayed in
figure~\ref{fig:boxdiags}, where $q=d$ corresponds to the case of
$B_d^0-\overline{B}_d^0$ and $q=s$ to $B_s^0-\overline{B}_s^0$ mixing. Due to
the hierarchy of the CKM matrix and the large top-quark mass, the short-distance
contribution is top and CKM enhanced, whilst the long-distance contribution is
CKM suppressed, so that the short-distance contribution dominates. This feature
makes lattice QCD computations of the corresponding matrix elements feasible (as
opposed to e.g. $D^0$-$\overline{D}^0$ mixing, where lattice QCD predictions are
limited to the sub-leading short distance contribution,
see for example ref.~\cite{Bazavov:2017weg}).

There are 5 independent, parity even, local, dimension 6 operators $\mc{O}_i$ in
an effective Hamiltonian for $\Delta B=2$. In the SM, only the operator
$\mc{O}_1$ contributes to the mass difference $\Delta m_q$, whilst the operators
$\mc{O}_1$, $\mc{O}_2$ and $\mc{O}_3$ contribute to the width difference. The
operators $\mc{O}_4$ and $\mc{O}_5$ do not contribute in the Standard Model, but
by virtue of being a loop mediated process, neutral meson mixing is a sensitive
probe for New Physics. To account for possible beyond the Standard Model
scenarios, a precise determination of the non-perturbative matrix elements of
all five operators is desirable.

The mass differences $\Delta m_q$ can be parameterised in terms of known
functions, CKM factors and non-perturbative matrix elements which can be
computed via lattice QCD computations. More precisely
\begin{equation}
  \Delta m_q = \frac{G_F^2 m_W^2 m_{B_q}}{6\pi^2} \abs{V_{tb}V_{tq}^*}^2 S_0(x_t) \eta_{2B} f_{B_q}^2 \hat{B}^{(1)}_{B_q}\,,
\end{equation}
where $q=d$ or $s$, $x_t = m_t^2/M_W^2$, $S_0(x_t)$ is an Inami-Lim
function~\cite{Inami:1980fz} and $\eta_{2B}$ captures short-distance QCD
corrections. Furthermore $f_{B_q}$ is the decay constant of the $B_q$ meson and
$\hat{B}^{(1)}_{B_q}$ is the renormalisation group independent bag parameter for
the operator $\mc{O}^{(1)}$. Precise non-perturbative predictions of
$f_{B_{(q)}}^2 \hat{B}^{(1)}_{B_q}$ , attainable via lattice QCD simulations,
enable the extraction of the combination of CKM matrix elements
$\abs{V_{tb}V_{tq}^*}$. Additionally, it is convenient to define the $SU(3)$
breaking ratio $\xi$~\cite{Bernard:1998dg}, defined by
\begin{equation}
  \xi^2 = \frac{f^2_{B_s} \hat{B}^{(1)}_{B_s}}{f^2_{B_d} \hat{B}^{(1)}_{B_d}}\,.
\end{equation}
By combining $\xi$ with the experimental measurements of $\Delta m_d$ and
$\Delta m_s$, one can extract the ratio $\abs{V_{td}/V_{ts}}$ from
\begin{equation}
  \abs{\frac{V_{td}}{V_{ts}}}^2 = \left(\frac{\Delta m_d}{\Delta
    m_s}\right)_\mathrm{exp} \frac{m_{B_s}}{m_{B_d}}
  \left(\xi^2\right)_\mathrm{lat}\,.
\end{equation}
This is favourable since various systematic and part of the statistical
uncertainties cancel in such a ratio. The results for $\abs{V_{td}}$,
$\abs{V_{ts}}$, and their ratio provide constraints which enter the global fits
produced by the UT fit~\cite{UTfit:2006vpt} and CKM-fitter~\cite{Charles:2004jd}
groups.

\section{Neutral meson mixing results from lattice QCD}
The large bottom quark mass poses significant challenges for the direct
simulation of heavy-light systems. Currently all lattice simulations employ one
of the following two approaches. In the first approach the heavy quark is
simulated via an effective action such as static quarks, the non-relativistic
QCD action~\cite{Lepage:1992tx,HPQCD:2011qwj} or the Fermilab
method~\cite{Sheikholeslami:1985ij,El-Khadra:1996wdx}. In the second approach
simulations take place at unphysically light heavy quark masses. The results are
then extrapolated to the physical $b$-quark
mass~\cite{ETM:2013jap,Boyle:2018knm}. In addition, all current lattice QCD
results for neutral meson mixing include heavier-than-physical pion mass
ensembles. One leading uncertainty in current lattice QCD predictions stems from
the chiral-continuum extrapolation. This extrapolation depends strongly on the
available gauge field ensembles at or near the physical pion mass and gauge
field ensembles with fine lattice spacings.

\begin{figure}
  \centering
  \includegraphics[width=.45\textwidth]{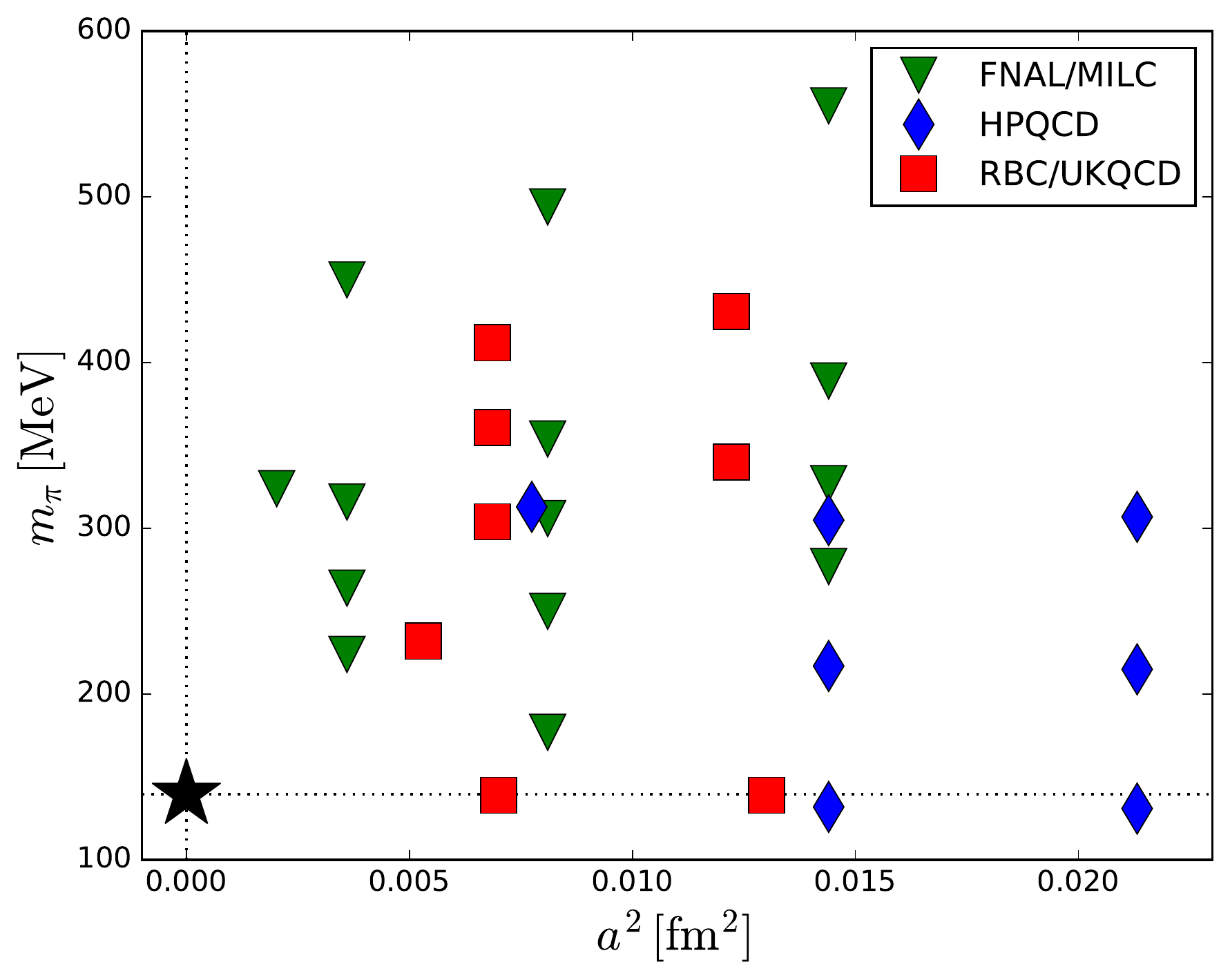}
  \includegraphics[width=.45\textwidth]{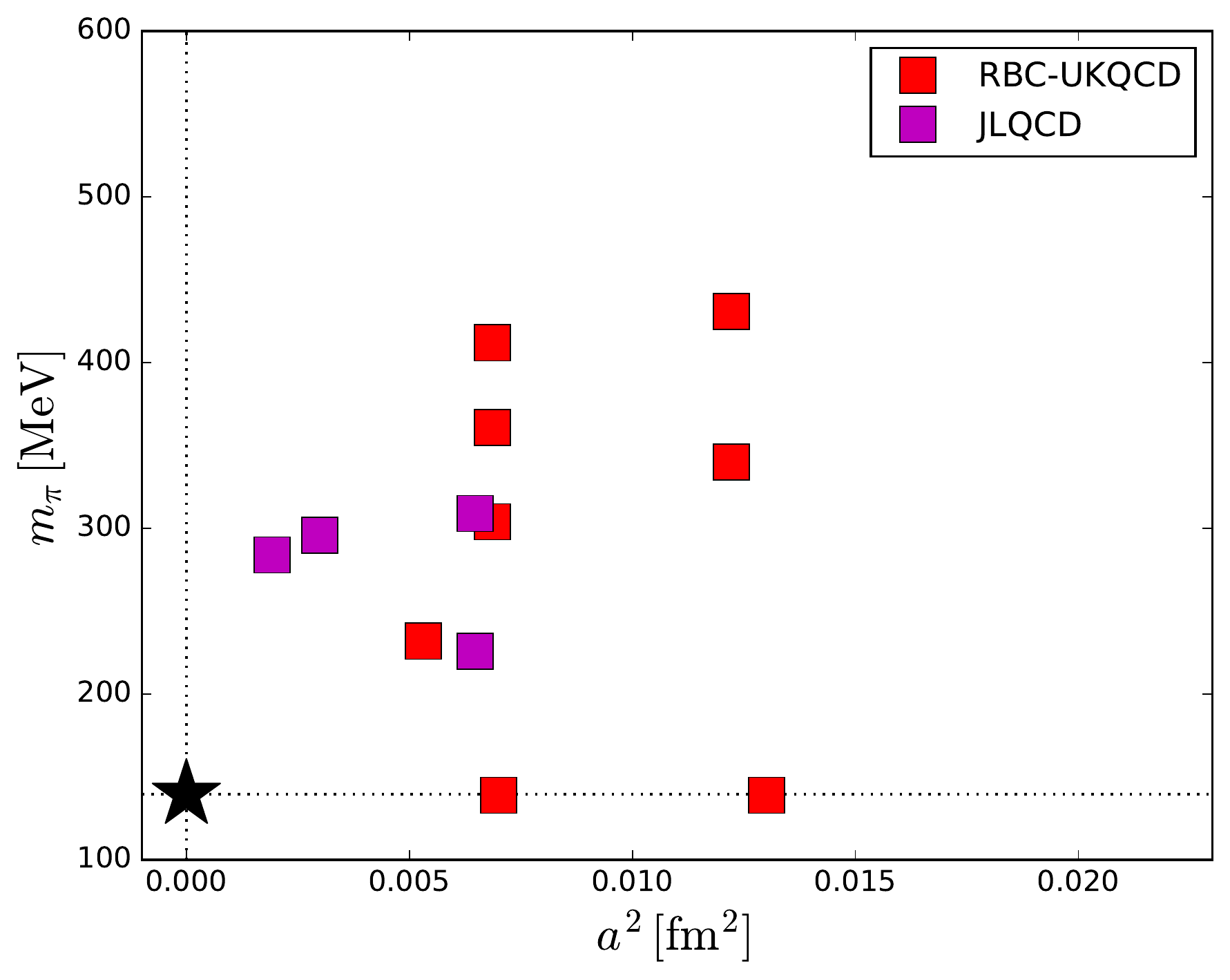}
  \caption{Gauge field ensembles as a function of the lattice spacing squared
    ($a^2$) and the pion mass squared $m_\pi^2$ used in recent computations of
    neutral meson mixing parameters. The black star corresponds to the physical
    point to which results need to be extrapolated. \emph{Left}: The ensembles
    entering the FNAL/MILC analysis~\cite{FermilabLattice:2016ipl}, the HPQCD
    analysis~\cite{Dowdall:2019bea} and the RBC/UKQCD
    analysis~\cite{Boyle:2018knm} are shown as green triangles, blue diamond and
    red squares, respectively. \emph{Right}: In Ref.~\cite{Boyle:2021kqn} the
    RBC/UKQCD dataset (red squares) is supplemented with a recently generated
    JLQCD dataset (magenta squares). The gauge fields ensembles used in
    refs~\cite{FermilabLattice:2016ipl,Boyle:2018knm,Boyle:2021kqn} include the
    dynamical effects of two degenerate light quarks and the strange quark
    ($N_f=2+1$), whilst the ensembles used in ref.~\cite{Dowdall:2019bea} also
    includes dynamical charm quark effects ($N_f=2+1+1$).}
  \label{fig:landscape}
\end{figure}

Figure~\ref{fig:landscape} presents an overview over the lattice spacing and
pion mass properties of the the gauge field ensembles that are used in the most
recent computations~\cite{FermilabLattice:2016ipl, Boyle:2018knm,
  Dowdall:2019bea, Boyle:2021kqn}.  For the discussion of older
results~\cite{Gamiz:2009ku,ETM:2013jap,Aoki:2014nga}, we refer the reader to the
recent update of the flavour lattice averaging group (FLAG)~\cite{Aoki:2021kgd}.

The Fermilab/MILC result~\cite{FermilabLattice:2016ipl} has been discussed in
previous iterations of this conference, so we only briefly summarise the
computation. The light quarks are discretised using the asqtad
action~\cite{Orginos:1999cr}, the $b$-quark is discretised using the Fermilab
method~\cite{Sheikholeslami:1985ij,El-Khadra:1996wdx}. The authors compute the
product $f_{B_q} \sqrt{\hat{B}^{(i)}_{B_q}}$ for $q=d,s$ and for
$i=1,...,5$. The bag parameters $\hat{B}^{(i)}_{B_q}$ are then found by taking
external input for $f_{B_q}$. The relative uncertainties for $f^2_{B_q}
\hat{B}_{B_q}^{(1)}$ are approximately 8 and 6\% for $q=d,s$, respectively. The
relative uncertainty of $\xi$ is approximately 1.5\%.

The recent HPQCD result~\cite{Dowdall:2019bea} uses the HISQ
action~\cite{Follana:2006rc} for the light quarks and the NRQCD action for the
$b$-quark~\cite{Lepage:1992tx,HPQCD:2011qwj}. The calculation provides values
for the bag parameters $\hat{B}_{B_q}^{(i)}$ for $i=1,...,5$ and $q=d,s$. The
decay constants are then taken as an external input to convert this to $f_{B_q}
\sqrt{\hat{B}_{B_q}^{(i)}}$. The inclusion of two ensembles at the physical pion
mass significantly reduces uncertainties associated to the chiral
extrapolation. The dominant uncertainty in this work arises from matching terms
of the order $\alpha_s^2$ and $\alpha_s\Lambda_\mathrm{QCD}/m_b$. The quoted
uncertainties for $\hat{B}^{(1)}_{B_q}$ are 5.0 and 4.3\% for $q=d,s$,
respectively, whilst the uncertainty of their ratio is 2.5\%. For $\xi$ an
uncertainty of 1.3\% is quoted.

The recent work by RBC/UKQCD~\cite{Boyle:2018knm} follows a different approach
in the treatment of the heavy quark. Here the (to a good approximation) chirally
symmetric domain wall fermion
action~\cite{Kaplan:1992bt,Blum:1996jf,Shamir:1993zy,Brower:2012vk} is used for
light and heavy quarks, albeit with different choices for the domain wall
parameters for the light and the heavy quarks~\cite{Boyle:2015kyy}. The heavy
quarks are simulated in the region from below the physical charm quark mass up
to approximately half the bottom quark mass. Ref.~\cite{Boyle:2018knm} provides
results for the ratios $\hat{B}_{B_s}/\hat{B}_{B_d}$, $f_{B_s}/f_{B_d}$ and
$\xi$. Because these ratios have a favourable behaviour as a function of the
heavy quark mass, they allow a controlled extrapolation to the physical
$b$-quark mass. As in the case of the HPQCD result, the inclusion of two
ensembles with physical pion masses removes most uncertainties associated to the
chiral extrapolation. The dominant uncertainty in this work is the extrapolation
to the physical $b$-quark mass. However, by addition of ensembles with smaller
lattice spacings, this uncertainty can be systematically improved upon. The
total uncertainty for the quantity $\xi$ is quoted at the percent level.

\section{Summary and future prospects}
The three computations~\cite{FermilabLattice:2016ipl, Dowdall:2019bea,
  Boyle:2018knm} highlighted above are based on completely disjoint gauge field
ensembles and are therefore independent predictions. These results are highly
complementary as all the discretisation of the light and the heavy quark action
differ between all three works. When comparing the numerical values obtained for
the bag parameters $B_{B_q}^{(i)}$, the product $f^2_{B_q} \hat{B}_{B_q}^{(1)}$,
and $\xi$, agreement between computations from the different groups can be seen.
Combining the lattice results with the experimentally measured mass differences
allows the extraction of $\abs{V_{td}}$, $\abs{V_{ts}}$ and
$\abs{V_{td}/V_{ts}}$ at the 2.6\%, 2.2\% and 1.3\%-level. However, in the
$\abs{V_{td}}-\abs{V_{ts}}$ plane, an approximately 2 sigma deviation remains
between the determinations from the different lattice results (see for example
figure~7 of ref.~\cite{Wingate:2021ycr}). Whilst the recent results are a
significant improvement, the uncertainties on the CKM matrix elements are still
dominated by the theory inputs, so additional work is required to improve upon
their accuracy.

Recently, work in progress was reported for a joint effort between the RBC/UKQCD
and the JLQCD collaborations~\cite{Boyle:2021kqn}. This supplements the existing
dataset from ref.~\cite{Boyle:2018knm} with additional ensembles with finer
lattice spacings provided by the JLQCD collaboration. The bound of only
simulating up to approximately half the physical $b$-quark mass was set due to
the lattice spacing on the finest ensemble~\cite{Boyle:2017jwu}. This can be
significantly extended by including the JLQCD ensembles, enabling simulations
close to the physical $b$-quark mass. The choice of the chirally symmetric
all-domain-wall fermion set-up significantly simplifies the renormalisation
structure and allows to draw on the expertise of the similar neutral kaon mixing
programme~\cite{Boyle:2017skn,Boyle:2017ssm}. Using this the authors anticipate
results for the full operator basis for the $B^0_d$ and the $B^0_s$ system in
addition to the $SU(3)$-breaking ratios presented in ref.~\cite{Boyle:2021kqn}.

This review focusses on the determination of the dimension 6 operators from
lattice QCD. Beyond this, in a recent computation~\cite{Davies:2019gnp} the
HPQCD collaboration presented the first lattice QCD computation of the matrix
elements of the dimension 7 operators which contribute to the width difference
at next to leading order. Finally since the last iteration a new computation of
the dimension 6 operators from sum rules has appeared~\cite{King:2019lal}.

\section*{Acknowledgements \label{sec:acknowledgements}}
J.T.T. would like to thank his collaborators, in particular M.~Della~Morte,
F.~Erben, J.~Flynn and O.~Witzel for critical proofreading of the draft. The
project leading to this application has received funding from the European
Union's Horizon 2020 research and innovation programme under the Marie
Sk{\l}odowska-Curie grant agreement No 894103.

\bibliographystyle{JHEP-notitle}
\bibliography{bibliography.bib}

\end{document}